\documentclass[a4paper,conference]{IEEEtran}

\usepackage{amsmath,amsfonts,color,comment,cite}
\usepackage[pdftex]{graphicx}

\newcommand{\EE}{\mathbb{E}}
\newcommand{\PP}{\mathbb{P}}

\newcommand{\ee}{\mathrm{e}}
\newcommand{\given}{\mid}

\newcommand{\J}{\mathcal{J}}
\newcommand{\K}{\mathcal{K}}
\newcommand{\LL}{\mathcal{L}}
\renewcommand{\hat}{\widehat}

\newcommand{\comp}{\mathsf{c}}
\newcommand{\mat}[1]{\mathsf{#1}}
\renewcommand{\vec}[1]{\mathbf{#1}}
\newcommand{\DD}{{\sc DD} }     %
\newcommand{\COMP}{{\sc COMP} } %
\newcommand{\SSS}{{\sc SSS} }   %

\newtheorem{theorem}{Theorem}
\newtheorem{lemma}{Lemma}

\begin{document}

\title{Improved Group Testing Rates \\ with Constant Column Weight Designs}

\author{\IEEEauthorblockN{Matthew Aldridge}
\IEEEauthorblockA{Department of Mathematical Sciences\\
University of Bath\\
Bath, BA2 7AY, UK \\     %
Email: m.aldridge@bath.ac.uk}
\and
\IEEEauthorblockN{Oliver Johnson}
\IEEEauthorblockA{School of Mathematics\\
University of Bristol\\
Bristol, BS8 1TW, UK \\
Email: maotj@bristol.ac.uk}
\and
\IEEEauthorblockN{Jonathan Scarlett}
\IEEEauthorblockA{Laboratory for Information and Inference Systems\\ \'Ecole Polytechnique F\'ed\'erale de Lausanne \\
CH-1015, Switzerland \\ 
Email: jonathan.scarlett@epfl.ch}}

\maketitle

\begin{abstract}
We consider nonadaptive group testing where each item is placed in a constant number of tests. The tests are chosen uniformly at random with replacement, so the testing matrix has (almost) constant column weights. We show that performance is improved compared to Bernoulli designs, where each item is placed in each test independently with a fixed probability. In particular, we show that the rate of the practical COMP detection algorithm is increased by 31\% in all sparsity regimes. In dense cases, this beats the best possible algorithm with Bernoulli tests, and in sparse cases is the best proven performance of any practical algorithm. We also give an algorithm-independent upper bound for the constant column weight case; for dense cases this is again a 31\% increase over the analogous Bernoulli result.
\end{abstract}

\vspace*{-2ex}
\section{Introduction}

The group testing problem, introduced by Dorfman \cite{dorfman}, is described as follows. Suppose we have a number of items, some of which are defective, and carry out a series of tests on subsets of items (`pools'). In the standard noiseless model we consider here, the result of a test is positive if the pool contains at least one defective item, and is negative otherwise. The task is to detect which items are defective using as few tests as possible, using only the list of testing pools and the outcomes of the corresponding tests.

We let $N$ denote the total number of items, let $K$ denote the number of defective items, and focus on the regime $K = o(N)$.  We suppose that the set $\K$ of defective items is chosen uniformly at random from the $\binom{N}{K}$ possible defective sets.

Finding the true defective set requires us to learn $\log_2 \binom{N}{K}$ bits of information. If an algorithm uses $T$ tests, then, following \cite{BJA}, we can consider the number of bits of information learned per test $\log_2 \binom{N}{K}/T$ as the \emph{rate} of the algorithm, and consider the \emph{capacity} as the supremum of all rates that can be achieved by any algorithm. We consider performance in the regime where the number of defectives scales as $K = \Theta(N^{\theta})$ for some density parameter $\theta \in (0,1)$.

In the adaptive case -- where we  choose successive testing pools using the outcomes of previous tests -- Hwang's generalized binary splitting algorithm \cite{hwang} shows that the capacity is $1$ for all $\theta \in (0,1)$ \cite{BJA}.
In this paper, we consider the nonadaptive case, where the testing pools are chosen in advance. %

It will be useful to list the testing pools in a binary matrix $\mat{X} \in \{0,1\}^{T\times N}$, where $x_{ti} = 1$ denotes that item $i$ is included in test $t$, with $x_{ti} = 0$ otherwise. Hence, the rows of the matrix correspond to tests, and the columns correspond to items.

We observe the outcomes $\vec y = (y_t) \in \{0,1\}^T$. A positive outcome $y_t = 1$ occurs if there exists a defective item in that test; that is, if for some $i \in \K$ we have $x_{ti} = 1$. A negative outcome $y_t = 0$ occurs otherwise.

As described in \cite{du}, it appears to be difficult to design a matrix  with order-optimal performance using combinatorial constructions.  Hence, a great deal of recent work on nonadaptive group testing has considered random design matrices, with a particular focus on Bernoulli random designs, in which each item is placed in each test independently with a given probability; see for example \cite{aldridge3, aldridge4, ABJ, scarlett, scarlett2, chan, atia}. The capacity for Bernoulli nonadaptive testing is
\begin{equation}
  	C(\theta) = \max_{\nu > 0} \min \left\{ \frac{\nu\mathrm{e}^{-\nu}}{\ln 2} \frac{1-\theta}{\theta}, \,h(\mathrm e^{-\nu}) \right\} , \label{eq:bern_rate}
\end{equation}
in particular yielding a capacity of $1$ for $\theta \leq 1/3$.  The achievability part of this result was given by Scarlett and Cevher \cite{scarlett}, and the converse by Aldridge \cite{aldridge4}.

This paper shows that improvements are possible with a different class of test designs, where each item is placed in a fixed number $L$ of tests, with the tests chosen uniformly at random with replacement. That is, independently within each column of $\mat X$, $L$ random entries are selected uniformly at random with replacement
and set to $1$. We  refer to these as `constant column weight' designs -- although strictly speaking, since the sampling is with replacement, some columns will have weight slightly less than $L$. The results of this paper could also be derived by sampling without replacement, thus yielding strictly constant column weights, with each item in \emph{exactly} $L$ tests, but we find that considering replacement is more convenient and leads to shorter proofs.
It will be convenient to parametrise $L = \nu T/K$, and we will later see that $\nu = \ln 2 \approx 0.693$ optimises our bounds for all $\theta$. This choice of $\nu$ gives a $50:50$ chance of tests being positive or negative, in contrast to Bernoulli testing with $\theta > 1/3$, where the optimal procedure has on average more positive than negative tests.

\begin{figure} 
\centering
\includegraphics[width=0.46\textwidth]{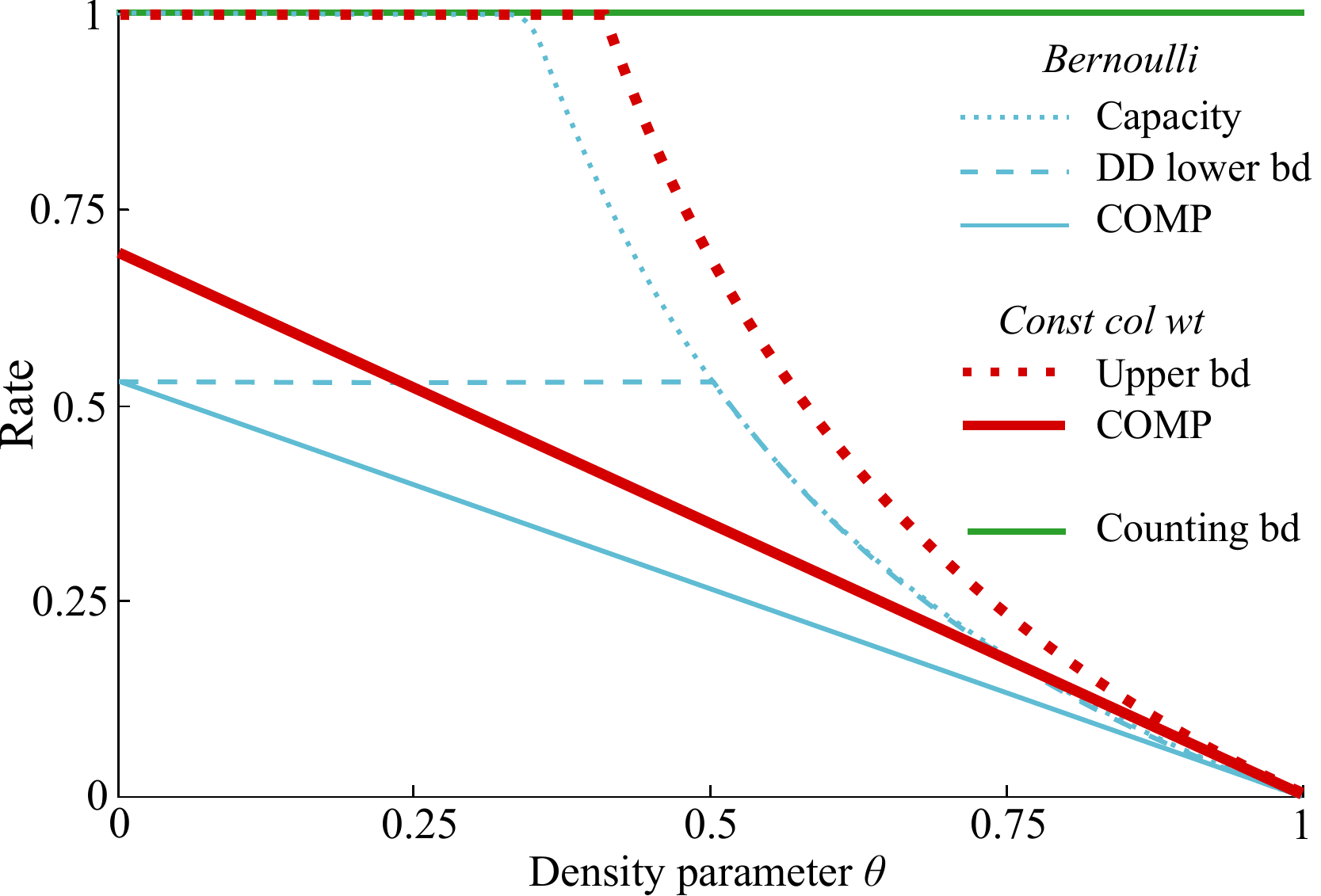}
\caption{Graph showing rates and bounds for group testing algorithms with Bernoulli designs and constant column weight designs.}
\label{figrate}
\vspace*{-1.5ex}
\end{figure}

The main result of this paper (Theorem \ref{compthm}) shows that with a constant column weight design, a simple and practical algorithm called COMP \cite{chan} %
achieves a rate of $0.693(1 - \theta)$. When used with a Bernoulli design, COMP has maximum rate $0.531(1 - \theta)$, so we  achieve a rate increase of $30.6\%$ for all $\theta$. Further, COMP with a constant column weight design outperforms \emph{any} algorithm used with a Bernoulli design for $\theta > 0.766$, and gives the best proven rate for a practical algorithm for $\theta < 0.234$ (beating a bound on Bernoulli designs with the DD algorithm \cite{ABJ}). These rates are shown in Fig.~\ref{figrate}.

In addition, we provide an
algorithm-independent converse (Theorem \ref{sssthm}); that is, an
upper bound on the rate that can be achieved by any detection algorithm with a constant column weight design. %
We conjecture that, as in \cite{aldridge4}, this converse is sharp and is
achieved by the \SSS algorithm.

We also give empirical evidence that, for a variety of algorithms,
constant column weight designs improve on Bernoulli designs in the finite blocklength regime (see Fig.~\ref{figsim}).

The idea of using constant column weight matrix designs is not a new one. The key  contribution of this paper is to rigorously analyse the performance of such designs, and to show that they can out-perform Bernoulli designs. %
We briefly mention some previous works that used constant column weight (and other related) designs.

M\'ezard \emph{et al.} \cite{mezard}, consider randomized designs with both fixed row and column weights, and with fixed column weights only.  It is noted therein that such designs can beat Bernoulli designs; however, they note that the analysis relies on a `short-loops' assumption that is shown to be rigorous only for $\theta > 5/6$, and in fact shown to be invalid for small $\theta$.  In contrast, we present analysis techniques that are rigorous for all $\theta \in (0,1)$. 

Kautz and Singleton \cite{kautz} observe that matrices corresponding to constant weight codes with a high minimum distance perform well in group testing. However, controlling the minimum distance is known to be a stringent requirement.

Wadayama \cite{wadayama} analyses constant row and column weight designs in the $K = cN$ regime, and demonstrates close-to-optimal asymptotic performance for certain ratios of parameter sizes. %

Chan \emph{et al.} \cite{chan} consider designs with constant row-weights only (like here, sampling with replacement) and find no improvement over Bernoulli designs.

The rest of this paper is organised as follows. Section \ref{simsec} demonstrates the empirical performance of the designs and algorithms discussed in this paper. Section \ref{ccsec} introduces some necessary results on the classical `coupon collector' problem. Section \ref{compsec} defines the \COMP algorithm, and proves our main theorem on the maximum rate for \COMP with constant column weight designs (Theorem \ref{compthm}). Section \ref{ssssec} defines the \SSS algorithm and proves an algorithm-independent upper bound for constant column weight designs.

\section{Simulations} \label{simsec}

\begin{figure} 
\centering
\includegraphics[width=0.48\textwidth]{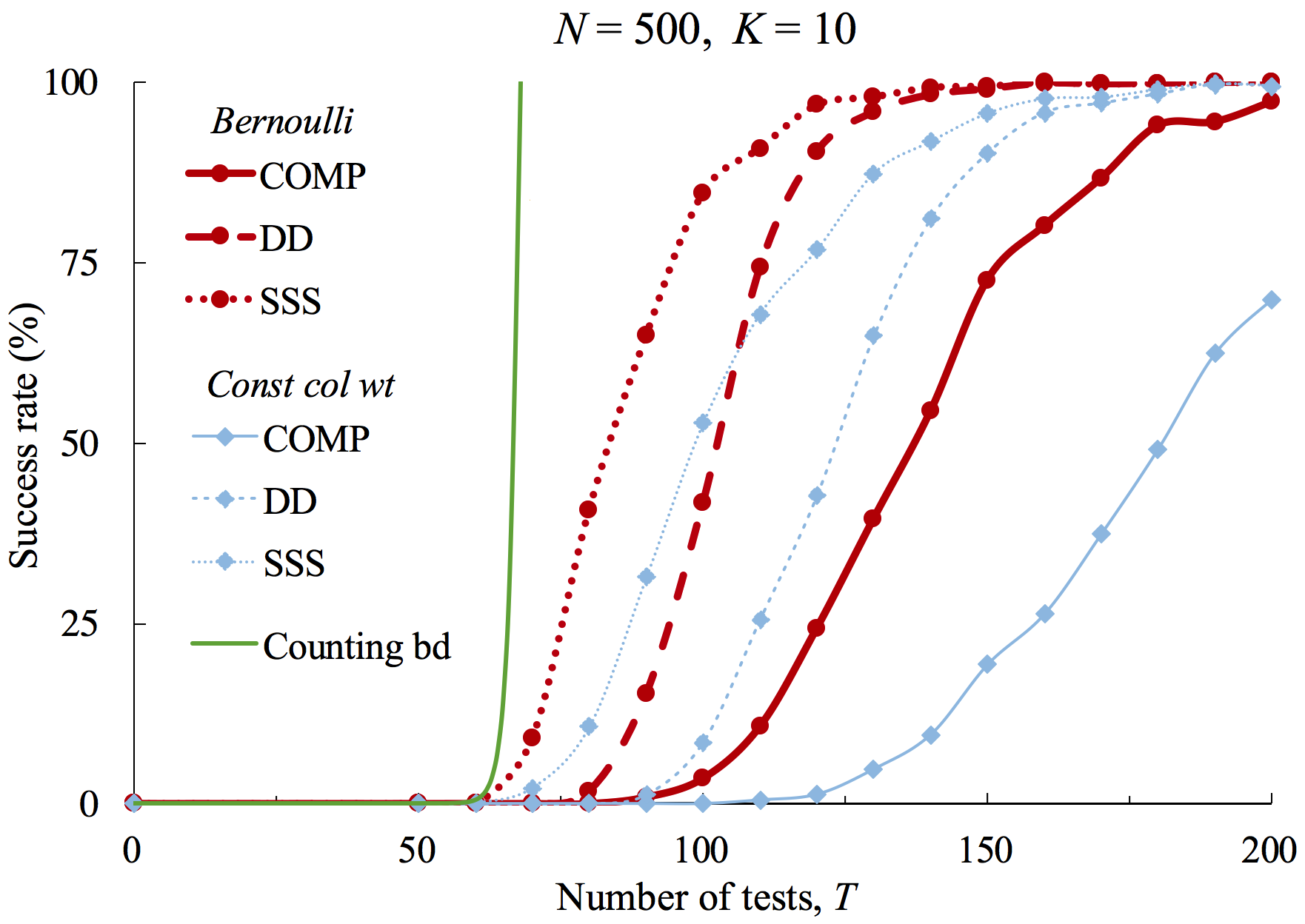}

\vspace{0.25cm}

\includegraphics[width=0.48\textwidth]{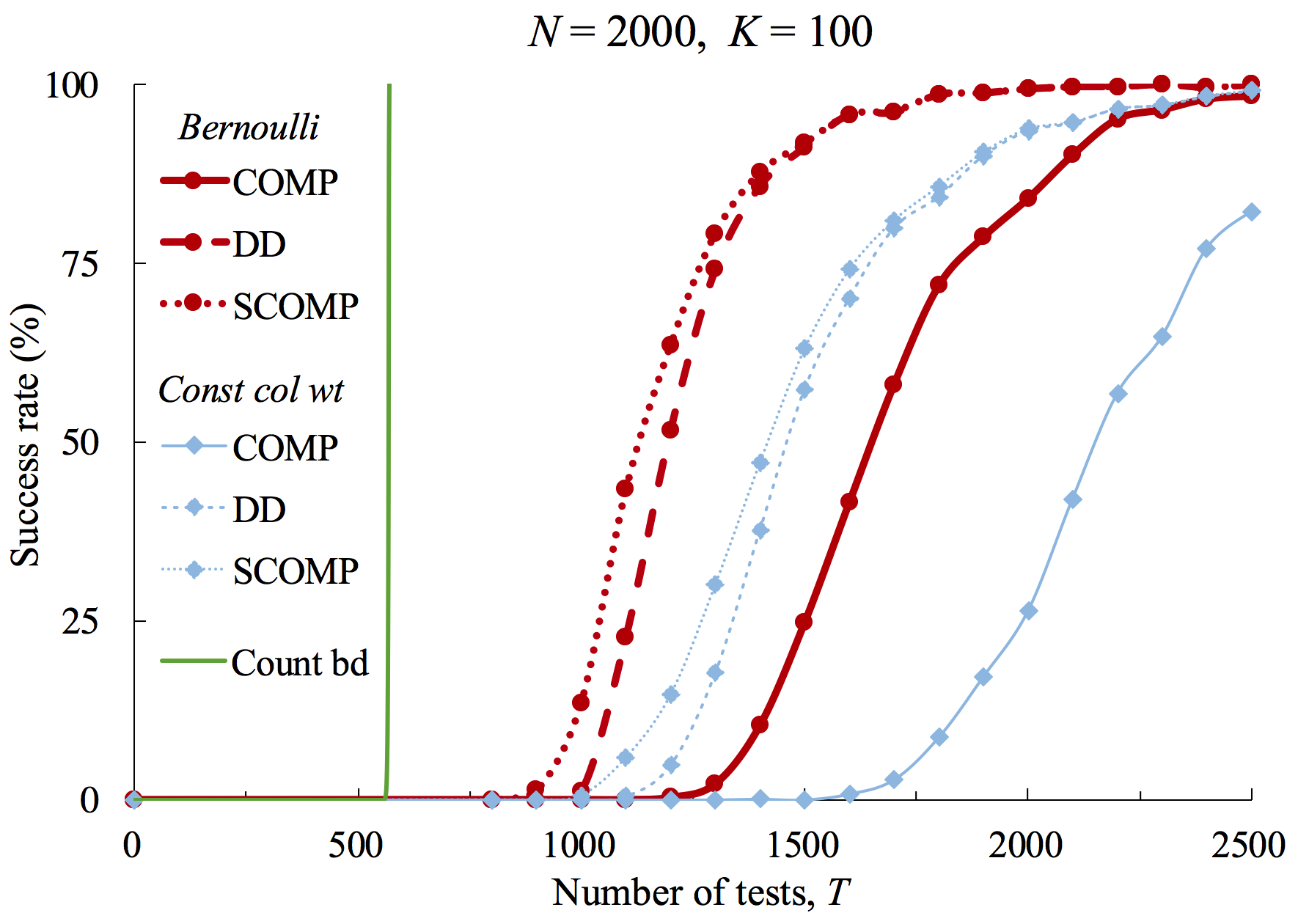}
\caption{Empirical performance results in the cases $N=500$, $K=10$, and $N = 2000$, $K = 100$, for a variety of algorithms with constant column weight and Bernoulli designs.  Each point represents $1000$ experiments.%
}
\label{figsim}
\vspace*{-1.5ex}
\end{figure}

Fig.~\ref{figsim} shows empirical performance results in an illustrative smaller, sparser case ($N=500$, $K=10=N^{0.371}$), and an illustrative larger denser case ($N = 2000$, $K = 100 = N^{0.606}$). In the first case, we show the COMP and SSS algorithms studied here, and the practical Definite Defectives (DD) algorithm \cite{ABJ}. In the second case, SSS is impractical, so we use the Sequential COMP (SCOMP) approximation \cite{ABJ}. Note that all algorithms perform better under constant column weight designs, and the simple DD algorithm with a constant column weight design usually outperforms more complicated algorithms with a Bernoulli design. We also plot the `counting bound' universal
converse (see \cite{BJA,johnson39}), which shows
that for any design (adaptive or nonadaptive), any algorithm has a success probability bounded above by $2^T/\binom{N}{K}$. 
In the first case of Fig.~\ref{figsim} the constant column weight \SSS algorithm has empirical performance
close to this universal upper bound.

\section{Coupon collector results} \label{ccsec}

Recall the coupon collector problem, where uniformly random selections, with replacement, are
made from a population of $T$ different coupons.  After making $c$ selections, the collector has collected some (random) number $W$ of \emph{distinct} coupons. Clearly $W$ can be as large as $c$ if all $c$ selections are different, but it can be less if there are some `repeated' coupons.
The following results  will be useful later.

\begin{lemma} \label{couplem}
Consider a population of $T$ coupons. A collector makes $c$ selections, and finds she has $W(c)$ distinct coupons.
  \begin{enumerate}
    \item We have
      \begin{equation} \label{cce}
        \EE W(c) = \left(1 - \left(1 - \frac1T\right)^c\right) T ,
      \end{equation}
      and further, if $c = \alpha T$ for some $\alpha \in (0,1)$, then
        \[ \EE W(\alpha T) \sim (1 - \ee^{-\alpha})T \]
      as $T \to \infty$ (where here and subsequently $\sim$ denotes equality up to a multiplicative $1+o(1)$ term).
    \item Again when $c = \alpha T$, we have concentration of $W$ about its mean, in that for any $\epsilon > 0$,
    \[
      \PP \left( \left| W(\alpha T) - (1 - \ee^{-\alpha}) T \right| \geq \epsilon T \right) 
        \leq 2 \exp \left(-\frac{\epsilon^2 T}{\alpha} \right),
    \]
      for $T$ sufficiently large.
  \end{enumerate}
\end{lemma}
\begin{IEEEproof}
For part 1, by linearity of expectation, we have
  \begin{align*}
    \EE W(c)
      &= \sum_{j = 1}^T \PP(\text{coupon $j$ in first $c$ selections}) \\
      &= \sum_{j = 1}^T \left(1 - \left(1 - \frac1T\right)^c \right) \\
      &= \left(1 - \left(1 - \frac1T\right)^c \right) T ,
  \end{align*}
as desired. The asymptotic form follows immediately.

For part 2, we use McDiarmid's inequality \cite{mcdiarmid}, which characterizes the concentration of functions of independent random variables when the bounded difference property is satisfied.  Write $Y_1, Y_2, \dots, Y_c$ for the labels of the selected coupons, and
  $W(c) = f(Y_1, Y_2, \dots, Y_c)$
for the number of distinct coupons. Note that here we have the bounded difference property, in that
  \[ \big| f(Y_1, \dots, Y_j, \dots, Y_c)
         - f(Y_1, \dots, \hat Y_j, \dots, Y_c) \big| \leq 1 \]
for any $j$, $Y_1, \dots, Y_c,$ and $\hat Y_j$, since the largest difference we can make is swapping a distinct coupon $Y_j$ for a non-distinct one $\hat Y_j$, or vice versa.
McDiarmid's inequality \cite{mcdiarmid} gives that
  \[
    \PP \big( \big| f(Y_1, \dots, Y_c)
      - \EE f(Y_1, \dots, Y_c) \big| \geq \delta \big)
      \leq 2 \exp \left( -\frac{2 \delta^2}{c} \right) .
  \]
Setting $\delta = \epsilon T$ and $c = \alpha T$ gives the desired result; we crudely remove the factor of $2$ from the exponent to account for the fact that we are considering deviations from the asymptotic mean instead of the true mean.
\end{IEEEproof}

\section{COMP} \label{compsec}

The COMP algorithm is due to Chan \emph{et al.} \cite{chan}. It is based on the observation that any item in a negative test is definitely non-defective; COMP then declares all remaining items to be defective. %
Hence, COMP succeeds if and only if each of the $N - K$ non-defectives appears in some negative test.

Chan \emph{et al.} \cite{chan} show that COMP with Bernoulli tests succeeds with $T > (1+\epsilon) \ee K \ln N$ tests, for a rate of $(1/e\ln 2)(1 - \theta) \approx 0.531(1 - \theta)$, and Aldridge \cite{aldridge4} showed that COMP can do no better than this with Bernoulli tests.

The following theorem reveals that constant weight columns improve the achievable rate by COMP.

\begin{theorem} \label{compthm}
Consider group testing with a constant column weight design and the COMP algorithm. Write
  \[ T^*_{\mathrm{COMP}} = \frac{1}{\ln 2}\, K \log_2 N . \]
Then, for any $\epsilon > 0$, with $T \geq (1 + \epsilon) T^*_{\mathrm{COMP}}$ tests, the error probability can be made arbitrarily small for $N$ sufficiently large, while with $T \leq (1 - \epsilon) T^*_{\mathrm{COMP}}$ tests, the error probability is bounded away from $0$.  Hence, the maximum achievable rate is $\ln 2\,(1 - \theta) \approx 0.693(1 - \theta)$.
\end{theorem}

Note that this shows that COMP with a constant column weight design outperforms any algorithm with a Bernoulli
design for large $\theta$, since it beats the rate in \eqref{eq:bern_rate} for $\theta > 1/(\ee (\ln 2)^2) \approx 0.766$.
Further, COMP improves the region where `practical' algorithms  work, in that it beats the best-known rate of a practical algorithm, previously given by the DD algorithm \cite{ABJ}, for $\theta < 1 - 1/(\ee \ln 2)^2 \approx 0.234$.

\begin{IEEEproof}
We start by following \cite[Remark 18]{ABJ}.  Recall that we write $L = \nu T/K$ for the %
weight of each column. We also write $M$ for the number of positive tests.
We know that COMP succeeds if and only if every non-defective item appears in a negative test. The probability that a given non-defective item appears in a negative test is $1$ minus the probability it appears only in positive tests, which, conditional on $M$, is $1 - (M/T)^L$. Note that since our design has independent columns, %
the events that particular non-defective items appear in a negative test are  conditionally independent given the  list of positive tests; this also holds for Bernoulli designs, but not for constant row-and-column designs as studied by Wadayama \cite{wadayama}.

Hence, given the number of positive tests,  COMP has
  \begin{equation} \label{comp1}
    \PP(\text{success} \mid M)
      = \left(1 - \left( \frac MT \right)^L \right)^{N - K} .
  \end{equation}
  
We start with achievability. For any $m^*$, we have
  \begin{align*} 
    \PP(\text{success})
      &= \PP(M  <   m^*) \PP(\text{success} \given M  <   m^*) \\
        &\qquad {}+\PP(M \geq m^*) \PP(\text{success} \given M \geq m^*) \\
      &\geq \PP(M \geq m^*) \PP(\text{success} \given M = m^*) ,
  \end{align*}
since the success probability \eqref{comp1} is decreasing in $M$.

The number of positive tests $M$ can be thought of as the number of distinct coupons collected from a population of $T$ (all tests) by $KL$ coupons: $L$ for each column in a total of $K$ columns corresponding to defective items.  Hence, by Lemma \ref{couplem}, we have concentration of $M$ about its mean $\EE M = (1 - \ee^{-\nu})T$. Hence, setting
  $m^* = (1 - \ee^{-\nu} - \epsilon)T$ ,
we have, for $N$ sufficiently large, 
  \begin{align*}
    \PP(\text{success})
      &\geq (1 - \epsilon) \PP\big(\text{success} \given
        M = (1 - \ee^{-\nu} - \epsilon)T\big) \\
      &= (1 - \epsilon)\left(1 - \left( \frac {(1 - \ee^{-\nu} - \epsilon)T}{T} \right)^L \right)^{N - K} \\
      &= (1 - \epsilon) \left(1 - \left(1 - \ee^{-\nu} - \epsilon \right)^{\nu T/K} \right)^{N - K} .
  \end{align*}

It is easy to check by differentiating that $(1 - \ee^{-\nu} - \epsilon)^\nu$ is maximised at $\ee^{-\nu} = (1 - \epsilon)/2$, which approaches $\nu = \ln 2$ for small $\epsilon$. We therefore set $\nu = \ln 2$, %
yielding
  \begin{align*}
    &\PP(\text{success}) \\
      &\ \geq (1 - \epsilon) \left(1 - \left(\frac12 - \epsilon\right)^{(\ln 2)T/K}\right)^{N-K} \\
      &\ \geq (1 - \epsilon) \left(1 - (N-K)\left(\frac12 - \epsilon\right)^{(\ln 2)T/K} \right) \\
      &\ = (1 - \epsilon) \left(1 - \exp\left( \frac{\ln 2}{K}\,T\,\ln(1/2 - \epsilon) + \ln(N-K)    \right)\right) .
  \end{align*}

We see that the success probability can be made arbitrarily close to $1$ by choosing $\epsilon$ sufficiently small, provided that
  \[ T \geq -(1 + \epsilon)\frac{K \ln(N-K)}{\ln 2 \ln(1/2 - \epsilon)} . \]
The achievability result follows on noting that
  \[ -\frac{\ln(N-K)}{\ln(1/2 - \epsilon)} \to -\frac{\ln N}{\ln(1/2)} = \log_2 N \]
as $N \to \infty$ and $\epsilon \to 0$, since $K = o(N)$.

We now turn to the converse. By a similar argument to the above, for any $m^*$, we have
  \begin{align} 
    \PP(\text{success})
      &= \PP(M  <   m^*) \PP(\text{success} \given M  <   m^*) \notag \\
        &\qquad {}+ \PP(M \geq m^*) \PP(\text{success} \given M \geq m^*) \notag \\
      &\leq \PP(\text{success} \given M = m^*) + \PP(M \geq m^*). \label{compcon}
  \end{align}
We now pick $m^* = (1 - \ee^{-\nu} + \epsilon)T$ .
By the concentration in Lemma \ref{couplem}, this gives that for any $\epsilon$ and for $N$ sufficiently large we have $\PP(M \geq m^*) \leq \epsilon$.
It remains to bound the first term of \eqref{compcon} %
by expanding out $(1 - ( 1 - \ee^{-\nu} + \epsilon )^L )^{N - K}$ one term further, using the inclusion--exclusion formula, then bounding as before. We omit full details due to space constraints.
\end{IEEEproof}

\section{Algorithm-independent converse} \label{ssssec}

In this section, we give an upper bound on the rate of group testing with constant weight columns that holds for any detection algorithm and any choice of $L$.

The proof, which is based on the capacity converse for Bernoulli testing \cite{aldridge4}, depends on analysis of a particular algorithm called SSS, which was studied in detail by Aldridge, Baldassini and Johnson \cite{ABJ}. The SSS algorithm declares as its estimate of $\K$ the \emph{smallest satisfying set} (if a unique such set exists). A set $\LL \subseteq \{1,2,\dots,N\}$ is \emph{satisfying} for the design $\mat X$ and outcomes $\vec y$ if using design $\mat X$ with true defective set $\LL$ would indeed give outcomes $\vec y$; in other words, the estimate $\LL$ `satisfies' the observations. The SSS algorithm chooses the satisfying set $\LL$ that minimises $|\LL|$, if a unique such $\LL$ exists, and declares an error (say) otherwise.

\begin{theorem} \label{sssthm}
Consider group testing with a constant column weight design and any detection algorithm. Write
  \begin{equation} \label{Tstar}
    T^* = \max \left\{ K \log_2 \frac NK\,, \ \frac{1}{\ln 2}\, K \log_2 K \right\} .
  \end{equation}
Then, for any $\epsilon > 0$, with $T \leq (1 - \epsilon) T^*$ the error probability is bounded away from $0$.

Hence, using a constant weight column design, the capacity is bounded above by
  \[ \min \left\{1,\  \ln 2\,\frac{1 - \theta}{\theta} \right\} \approx \min\left\{1,\  0.693\frac{1 - \theta}{\theta} \right\} . \]
\end{theorem}

\begin{IEEEproof}
We follow Aldridge's proof of a similar result for Bernoulli testing \cite{aldridge4}.
As shown there, it suffices to bound the error probabilities of the COMP and SSS algorithms. %
We have bounded the rate of COMP in Theorem \ref{compthm}, and it easy to see that the COMP bound satisfies this theorem, since
  \[ \frac{1}{\ln 2}K\log_2 N \geq T^* , \]
where $T^*$ is as in \eqref{Tstar}. Hence, it suffices to bound the error probability of the SSS algorithm, which we do below.

The first term in the maximum in \eqref{Tstar} is the counting bound, which holds for arbitrary designs \cite{BJA}. It remains to show the second term.

Following \cite{ABJ}, the error probability of SSS is bounded by
\begin{align}
	\mathbb P(\text{error}) 
    	&\geq \sum_{j=1}^k (-1)^{j+1} \sum_{|\J| = j} \PP(A_{\J}) \notag \\
    	&\geq \sum_{|\J| = 1} \PP(A_{\J}) 
          - \sum_{|\J| = 2} \PP(A_{\J}), \notag
\end{align}
where $A_{\J}$ is the event that none of the $T$ tests includes a %
defective item from $\J \subseteq \K$.

In coupon collector terms, $A_{\J}$ is the event that the $jL$ coupons of defective items from $\J$ only hit the tests already hit by the $(K-j)L$ coupons of the other defective items. 
Given $S_\J$, this number of `already hit' tests, the probability that the other $jL$ coupons only hit these tests is $(S_\J/T)^{jL}$.

Hence, we have, conditional on the $S_\J$s, that
  \begin{align}
    \PP(\text{error} \given \{S_\J\})
    	\geq \sum_{|\J| = 1} \left( \frac{S_\J}{T} \right)^{jL}
          - \sum_{|\J| = 2} \left( \frac{S_\J}{T} \right)^{jL}.
          \label{SJbound}
  \end{align}

We again parametrise $L$ as $L = \nu T / K$ for some (arbitrary) $\nu > 0$. From Lemma \ref{couplem}, the number of tests $S_\J$ already hit is exponentially concentrated about its mean, which behaves as
  $\EE S_\J \sim (1 - \ee^{-\nu(1 - j/K)}) T$.
Further, for $j = 1,2$, this simplifies to
  $\EE S_\J \sim (1 - \ee^{-\nu}) T$.
  
We shall condition on the $S_\J$ being suitably close to their means for $|\J| = 1,2$. Specifically, we define the event
  \[ B = \left\{ \begin{split}
       S_\J &\geq (1 - \ee^{-\nu} - \epsilon) T
         \quad \text{for all $|\J| = 1$} \\
       S_\J &\leq (1 - \ee^{-\nu} + \epsilon) T
         \quad \text{for all $|\J| = 2$} \end{split} \right\} . \]
Then by Lemma \ref{couplem} and the union bound,
  \[ \PP(B) \geq 1 - \left(K + \binom K2 \right)
                 2 \exp \left( - \frac{\epsilon^2 KT}{\nu} \right)
            \geq 1 - \epsilon  \]
for $T$ sufficiently large.

Using a similar argument to Section \ref{compsec}, we have
  \begin{align*}
    \PP(\text{error})
      &= \PP(B^\comp) \, \PP(\text{error} \given B^\comp) 
           + \PP(B) \,\PP(\text{error} \given B) \\
      &\geq \PP(B) \,\PP(\text{error} \given B) \\
      &\geq (1 - \epsilon)\,\PP(\text{error} \given B) 
  \end{align*}
for $N$ sufficiently large.

Combining the definition of the event $B$ with \eqref{SJbound}, we have
  \begin{align*}
    &\PP(\text{error} \given B) \\
      &\  \geq \sum_{|\J|=1} \left(\frac{\EE(S_\J\given B)}{T}\right)^{jL}
          - \sum_{|\J|=2} \left(\frac{\EE(S_\J\given B)}{T}\right)^{jL} \\
      &\  \geq \sum_{|\J|=1} \left(\frac{(1 - \ee^{-\nu} - \epsilon)T}{T}\right)^{jL}
           - \sum_{|\J|=2} \left(\frac{(1 - \ee^{-\nu} + \epsilon)T}{T}\right)^{jL} \\
      &\  =K (1 - e^{-\nu} - \epsilon)^L
             - \frac12 K^2 (1 - e^{-\nu}+\epsilon)^{2L} \\
      &\  = K (1 - e^{-\nu} - \epsilon)^{\nu T/K}
             - \frac12 K^2 (1 - e^{-\nu}+\epsilon)^{2\nu T/K} \\
      &\  = K(1 - e^{-\nu} - \epsilon)^{\nu T/K}
        \left( 1 - \frac{K}{2} \left( \frac{(1 - e^{-\nu} + \epsilon)^2}{1 - e^{-\nu} - \epsilon}\right)^{\nu T/K}  \right).
  \end{align*}
  
For $\epsilon$ sufficiently small, this is, like before, minimized arbitrarily close to $\nu = \ln 2$. Also for $\epsilon$ sufficiently small,
 \[ \frac{(1 - e^{-\nu} + \epsilon)^2}{1 - e^{-\nu} - \epsilon} \]
is arbitrarily close to $1 - e^{-\nu}$. Hence, using $\nu = \ln 2 + o(1)$, %
we have, as $\epsilon \to 0$,
  \begin{multline*}
    \PP(\text{error} \given B) 
    \geq \big(1 - o(1)\big) K(1 - e^{-\ln 2})^{(\ln 2) T/K} \\
        \times \left( 1 - \frac{1}{2} K 
        \big(1 - e^{-\ln 2}\big)^{(\ln 2) T/K} 
        \right),
  \end{multline*}
where we have absorbed all $o(1)$s into the first term. But
  \begin{align*}
    K (1 - \ee^{-\ln 2})^{(\ln 2) T/K}
      &= K \left( \frac12 \right)^{(\ln 2) T/K} \\
      &= 2^{\,\log_2 K - (\ln 2) T/K} ,
 \end{align*}
and since the error probability is decreasing in $T$, we have for
  \[ T \leq (1-\epsilon) \frac{1}{\ln 2} K \log_2 K . \]
that the error probability for SSS is bounded away from $0$.
\end{IEEEproof}

\section{Conclusions and future work}

We have shown that constant column weight matrix designs, instead of Bernoulli designs, can improve the rate that can be proved for the simple, yet order optimal,
\COMP algorithm. Further, we have shown an algorithm-independent converse, giving an upper bound
on the rate which can be achieved by any algorithm under this design. We conjecture
that this converse is sharp, in the sense that this performance can 
asymptotically be achieved
by the \SSS algorithm, and even by the simpler \DD algorithm for certain parameter
values. We hope to adapt the techniques of \cite{ABJ, scarlett} to resolve this conjecture 
in future work.

\section*{Acknowledgments}

M.~Aldridge was supported by the Heilbronn Institute for Mathematical Research.  J.~Scarlett was supported by the `EPFL fellows' programme (Horizon2020 grant 665667).

\end{document}